\newif\iffiginclude
\figincludefalse        
\iffiginclude
\documentstyle[prl,aps,preprint,tighten,psfig]{revtex}
\else
\documentstyle[prl,aps,preprint,tighten]{revtex}
\fi

\begin{document}

\preprint{\bf UPR--600--T/rev}



\title{ EXTREME DOMAIN WALL--BLACK HOLE COMPLEMENTARITY IN  $N=1$
 SUPERGRAVITY WITH A GENERAL DILATON COUPLING}

\author{MIRJAM CVETI\v C}

\address{Department of Physics, University of Pennsylvania \\
Philadelphia, PA 19104-6396, USA \\}

\maketitle

\begin{abstract}
We study  supersymmetric (extreme)  domain walls in four-dimensional ($4d$)
$N=1$  supergravity theories with a general dilaton coupling
$\alpha > 0$.   Type  I walls, which   are
static,  planar (say, in ($x,y$) plane)
configurations, interpolate between Minkowski space-time and  a vacuum  with a
varying dilaton field.  We classify
their global space-time  with respect to the
value of  the  coupling $\alpha$. $N=1$ supergravity with
 $\alpha =1$, an effective theory   from superstrings,  provides a
 dividing line between the theories  with $\alpha>1$, where there is
 a  naked (planar) singularity  on  one side of the  wall,
 and the theories  with $\alpha<1$, where the singularity of the
of the  wall is covered by the horizon.
The global space-time (in $(t,z$) direction) of the extreme walls  with
the  coupling $\alpha$
is the same as the global space-time  (in ($t,r)$ direction)
of the extreme magnetically charged black
holes  with the  coupling $1/\alpha$.
\end{abstract}

\vskip 10mm
\centerline{\it To be published in Physics Letters B}
\vskip 20mm

Over the last few years  topological defects in general, and  black
holes,\footnote{For a review see Ref. \cite{H} and references therein.} in
particular, have
been studied extensively  in theories  in which  an additional scalar field,
the
dilaton, couples to such defects.  The dilaton adds new features to the
nature of such configurations, in particular in the region of the space-time
where it blows-up. A special class of such  configurations correspond to the
extreme configurations,  which can be shown to  be  supersymmetric
configurations whose mass saturates the corresponding Bogomol'nyi bound.
They are thus of special interest since they correspond to
configurations  with the minimal energy in its class, and can be viewed as
geniune solitons.

In this paper we study   extreme domain wall configurations interpolating
between isolated  supersymmetric vacua  of the
four-dimensional ($4d$)  $N=1$ supergravity theory with a
general dilaton coupling specified by parameter $\alpha>0$. In particular we
 classify their   space-time structure with respect to the coupling $\alpha$.
 The dilaton
field, a scalar component of a linear multiplet, acts  as an additional
matter source, adding new features to configurations in this type of
supergravity theories. We shall
concentrate on  extreme configurations, {\it i.e.},   those are the
ones which  preserve part  ``$N={1\over 2}$'' of the supersymmetry and they
saturate the
corresponding  Bogomol'nyi bound for the energy density of the wall.
Specifically, we address extreme Type I dilatonic
 domain walls, which turn out to be   static, planar
configurations, interpolating between Minkowski space-time and  a vacuum  with
a
varying dilaton field. We shall compare their global space strucutre with the
one of the corresponding   extreme   charged dilatonic black
 holes\cite{GA,GM,GHS,HW,KLOPP,H}.


We study such configurations within
$4d$ $N=1$ supergravity  theory with  a linear super-multiplet
whose scalar component corresponds to the dilaton field.  Within the K\" ahler
super-space formalism\footnote{ The Lagrangian with
the chiral superfield  ${\cal S}$  is  at the classical level equivalent  to
the
one with the linear supermultiplet ${ L}$, where  ${\cal S}$
 and  ${ L}$ are related through the duality transformation. See, {\it
e.g.},  Ref.\cite{ABGG} and references therein.}
 the dilaton field
$\phi$ is represented as a real part of the scalar component
$ S \equiv {\rm e}^{-2\phi/\sqrt\alpha}  + ia\ $
of the  chiral superfield ${\cal S}$. ${\cal S}$ has the following
 restricted form of the  K\" ahler potential
\begin{equation}
K({\cal S},\overline {\cal S}) = -\alpha\,{\rm ln}({\cal S} + \overline
{\cal S})\ .\label{kahl}
\end{equation}
and  no superpotential ($W({\cal S}) = 0$). The field
$\phi$ is  defined  in $S$ so that its kinetic energy is normalized.  Note,
that in $4d$ $N=1$ supergravity  the value of  $\alpha$ in
(\ref{kahl}) is an arbitrary positive constant  with
 $\alpha=1$  corresponding  to an effective   theory  of superstring
vacua.
 ${\cal S}$ also couples
linearly  to  the  kinetic energy of the Yang-Mills  superfield ${\cal W}_{\mu
a}$,
{\it i.e.}, yielding a  term in the Lagrangian proportional to
\begin{equation}
{\cal S} \ {\hbox {tr}} \left( {\cal W}_{\mu a}{\cal W}^{\mu a}\right)\
.\label{wa}
\end{equation}
In addition, the theory contains
 chiral  superfields, ${\cal T}_i$,  with  K\" ahler potential  $\break
K_M({\cal T}_i,\overline
{\cal T}_i)$ and nonzero superpotential $W_M({\cal T}_i)$.

The tree-level low-energy action of $4d$ $N=1$ supergravity theory  with a
linear multiplet is therefore specified by a separable K\" ahler potential
$K = K_M ({\cal T}_i, \overline{\cal T}_i) + K({\cal S} ,
\overline{\cal S} )$,
superpotential $W=W_M({\cal T}_i)$, and the gauge coupling function
 $f_{a b}=\delta_{a b}{\cal S}$.

In the following we shall
study the extreme domain walls and compare their global space-time to the one
of the   charged  black holes
within the above  class of supergravity theories.
For the sake of simplicity we shall assume that  walls are formed
due to isolated minima of one matter field $T$ (a scalar component of
a chiral superfield), only. In addition, we  address Abelian  charged black
holes
with one $U(1)$ gauge group factor (Kac-Moody level $k=1$), only.
We   also set the axion field
$a$ , the imaginary part of $S$,  to zero.
The bosonic part of the tree level Lagrangian is then of the
 form:\footnote{ We are concerned with  classical
solutions of the  low-energy effective action.  Thus, we neither include
higher
derivative  terms, nor  possible   mixed K\"
ahler-Lorentz and mixed  K\" ahler-Gauge anomalies (see, {\it e.g.},  Ref.
\cite{CO} and references therein).
Eventually, such corrections should   be included in the full
treatment of the theory.}

\begin{eqnarray}
&{\cal {L}} =\sqrt{-g} [- {1\over 2} R + K_{T \overline{T}}
\partial _\mu T \partial^\mu \overline{T} +
 \partial_\mu \phi \partial^\mu \phi
 - 2^{-\alpha}{\rm e}^{2\sqrt{\alpha}\phi}\tilde V
-{1\over 2}{\rm e}^{-2\phi/\sqrt{\alpha}}F_{\mu\nu}F^{\mu\nu} ]\ .\label{lag}
\end{eqnarray}
where
\begin{equation}
\tilde V = {\rm e}^{K_M} [K^{T \overline{T}} |D_T W_M|^2 -
(3-\alpha)|W_M|^2 ]\label{pot}
\end{equation}
is the part of the potential, which  depends only on the matter field $T$.
Here $K^{T \overline{T}} \equiv (K_{T \overline
{T}})^{-1} \equiv (\partial_T \partial_{\overline {T}}K_M)^{-1}$ and $D_T
W_M \equiv {\rm e}^{-K_M}\partial_{T}({\rm e}^{K_M} W_M)$.
We use the metric convention $(+---)$ and  set the   gravitational  constant
 $\kappa\equiv 8\pi G=1$.

The extreme  wall solutions correspond to the  the case when
 gauge fields are turned off, {\it i.e.}, $F_{\mu\nu}=0$, and the matter
potential $\tilde V$ (Eq.(\ref{pot})) has two isolated
 supersymmetric ($D_TW_M(T)=0$) minima  with  $W_M(T)=0$ and  $W_M(T)\ne 0$,
respectively.
The corresponding  wall solution (Type I) is a static, planar configuration
(say, in the $(x,y)$ plane located at
$z\sim 0$).  It interpolates between,
 say, $z>0$, the
 supersymmetric vacuum  with $W_M(T)=0$, and  $z<0$, the supersymmetric vacuum
with $W_M(T)\ne 0$.
The metric
{\it Ansatz} for planar, static domain wall
solutions
\begin{equation}
ds^2 = A(z)(dt^2 - dz^2 - dx^2 - dy^2)\label{met}
\end{equation}
is conformally flat. The scalar fields $T(z)$, and $\phi(z)$  depend
on $z$, only.  Using a
technique of the generalized Israel-Nester-Witten  form, similar to the one
developed
for the study of supergravity walls without the dilaton field in
Ref. \cite{CGR}, one can show that  the field equations,
the self-dual Bogomol'nyi equations, are of the form:
\begin{eqnarray}
&0={\rm Im} (\partial _{z}T {{D_{T}W_{M}} \over {W_M}})\nonumber \\
& \partial_{z} T = - (2^{-\alpha}A{\rm e}^{2\sqrt\alpha\phi})^{1/2}
{\rm e}^{K_M /2} |W_M| K^{T \overline{T}} _M
{{D_{\overline{T}}\overline{W} _M}
\over {\overline{W} _M}} \nonumber \\
&\partial_{z} {\rm ln} A = 2 (2^{-\alpha}A{\rm e}^{2\sqrt\alpha \phi})^{1/2}
{\rm e}^{K_{M}/2}|W_{M}|\nonumber \\
&\partial _{z}\phi = - \sqrt\alpha(2^{-\alpha}A{\rm
e}^{2\sqrt\alpha\phi})^{1/2}
{\rm e}^{K_{M}/2}|W_{M}| \  .\label{boge}
\end{eqnarray}
The energy density of the wall, $\sigma$, saturates the Bogmoln'yi bound. For
a thin wall  with the boundary conditions $A(0)=1$ and
$\phi(0)=\phi_0$,  $\sigma$ is of the form:
\begin{equation}
\sigma= \sigma_{ext}\equiv 2^{1-{\alpha \over 2}}{\rm
e}^{\sqrt\alpha
\phi _{0}}|{\rm e}^{{K_M} \over 2}
W_M|_{z=0^-}\label{sig}
\end{equation}
 Here the subscript $0^-$ refers to the side of the wall
with $W_M(T)\ne 0$.

The first two equations in (\ref{boge}) describe the evolution of the
matter field $T = T(z)$ with $z$. In (\ref{boge})  the third equation for
 $A(z)$ and the fourth equation for $\phi(z)$  imply:
\begin{equation}
\tilde A(z)\equiv A(z){\rm e}^{2\phi(z)/\sqrt\alpha} =
{\rm e}^{2\phi_0/\sqrt\alpha}\ .\ \label{ap}
\end{equation}
The above equation is true {\it everywhere} in the domain wall
background.  It implies that there is a choice (\ref{ap}) of the frame
where  the  metric is {\it flat}.

The explicit form of
solutions in the thin wall approximation  is of the form:
\begin{eqnarray}
&A(z) = 1,\ \ \ \ \ \ \ \ \ \ \
\ \ \ \ \ \ \ \ \ \ \ \ \ \ \ \ \ \ \ \ \ \ \ \ \ \ \ \ \ \
  z >0\ ;\nonumber \\
&A(z)
= \left[1 - {1\over 2}(\alpha - 1)\sigma_{ext}|z|\right]^{2 \over {\alpha-1}}\
,
 \ \alpha\ne
1\  ,   \ z<0\ ;\nonumber \\
 &A(z)={\rm e}^{-\sigma_{ext}|z|}\ ,\ \ \ \ \ \ \ \ \
\ \ \ \ \ \ \ \ \ \ \ \ \ \ \alpha=1 \ , \ z<0\ ,
 \label{mets}
\end{eqnarray}
and the dilaton field satisfies Eq.(\ref{ap}).
On one side of the wall ($z>0$)
the space-time is Minkowski; both,  $A(z)$ and
$\phi$, are constant. However, on the other side of the wall ($z<0$), both,
$A(z)$ and $\phi(z)$, change its value.
There the curvature is  of the form:
\begin{eqnarray}
R&= {{3 (2 - \alpha)
\sigma_{ext}^2}\over 2}\left[1 - {1\over 2}(\alpha -
1)\sigma_{ext}|z|\right]^{-
{{2\alpha} \over {\alpha-1}}}\ ,
\ \alpha\ne 1\ ;\nonumber\\
R&={3\over 2}\sigma_{ext}^2{\rm e}^{-\sigma_{ext}|z|}\ ,\
\ \ \ \ \ \ \ \ \ \ \ \ \alpha=1\ . \label{cur} \end{eqnarray}

Such walls therefore act as ``windows'' from the
Minkowski space-time into the new type of  space-times with varying dilaton
field. The nature of these space-times depends  crucially
on the value of  parameter $\alpha$:

\begin{itemize}
\item{$\alpha=0$ corresponds to the case of ordinary  supergravity
 walls \cite{CGR,CG}, {\it i.e.},   the dilaton field  does not couple to
the matter fields (which form
the wall).   For $z<0$ the induced
space-time side is    anti-de Sitter. At $(t=\infty ,z=-\infty,)$ there is a
Cauchy horizon  with the zero surface gravity (${\cal K}\equiv
{1\over 2}\partial_z [{\rm
ln}A(z)]_{z=-\infty}=0 $)   and thus zero temperature ($T\equiv{\cal
K}/(2\pi)=0$) \cite{CDGS}. Geodesic
 extensions of the space-time were given in
 Refs. \cite{CDGS} and \cite{GIBBIII}.
The most symmetric geodesic extension  comprises of
a  system of  an infinite lattice
of semi-infinite Minkowski space-times  separated by an anti-de Sitter core.
The Carter-Penrose diagram
in the $(z,t)$ direction
is given in  Figure 1a.}
\item{
$\alpha< 1$ corresponds to the walls where curvature
blows up at $z=-\infty $.  On the other hand,   null geodesics reach
$z=-\infty$ in infinite  affine time $\tau\equiv \int^{-\infty} d\, z$,
{\it i.e.},  $z=-\infty$ corresponds to the  event horizon. Therefore the
planar singularity  at $z=-\infty$ is {\it null}, {\it i.e.},
it is covered by the event  horizon.
The associated temperature $T=0$.
The Carter-Penrose diagram
in the $(z,t)$ direction
is given in  Figure 2a.}
\item{
$\alpha=1$ corresponds to the case  of stringy dilatonic wall \cite{C}
of $4d$ tree-level effective string action.
The wall interpolates between the constant dilaton vacuum
(Minkowski space-time) and   the linear dilaton vacuum, {\it i.e.},
$\phi=\sigma_{ext}|z|/2$.  Note, that now the frame (\ref{ap}) with the flat
metric
is the sigma model frame  (string frame) of the string theory, {\it i.e.},
strings do not see the wall.
 At   $z=-\infty$, where  the curvature (and the dilaton) blow up,   the
metric has an event horizon as well \cite{C} .  Again, the
singularity is null, {\it i.e.}, it is covered by the event horizon,
 and the  global space-time is
 the same as the one for the walls with $\alpha< 1$
(see Figure 2a).  However, the temperature
associated with the horizon is now finite, {\it i.e.},
 $T=\sigma_{ext}/(4\pi)$.}
\item{
 $\alpha>1$ corresponds to the case where  the metric becomes singular
at  a {\it  finite} coordinate
distance $|z|_{sing} = {2\sigma_{ext}/ (\alpha - 1)}$ from the wall.\footnote{
Such a class of domain walls were first found \cite{CY} for  discrete values
of parameter $\alpha
=(2, 3, ...)$ .}
  For $\alpha \ne 2$, the curvature
blows-up  at $|z|_{sing}$. For $\alpha =2$, $R = 0$ but $R_{\mu \nu}R^{\mu \nu}
= \infty$ at the singularity.   Null geodesics
reach $|z|_{sing}$  is a {\it finite} time  {\it i.e.}, the affine time
$\tau$ is finite. Thus,
the planar singularity is naked.  The Carter-Penrose diagram
in the $(z,t)$ plane for this type of dilatonic
 domain walls  is given in  Figure 2a.  In addition, the surface gravity ${\cal
K}$ blows-up at this point, and thus $T=\infty$.}
\end{itemize}

Extreme stringy
dilatonic walls ($\alpha=1$, $T=\sigma_{ext}/(4\pi)$)
therefore serve as a dividing line between extreme walls ($\alpha<1$, $T=0$)
with  the (planar) singularity covered by the horizon  and the extreme
walls ($\alpha>1$, $T=\infty$) with the naked  (planar)
singularity.

We would now like to compare the above solutios to the ones of the
 extreme magnetically charged black
holes with  a general dilaton coupling \cite{GM,HW}.\footnote{Note that in
general,
 Lagrangian (\ref{lag}) describes
the bosonic part of  $4d$ $N=1$
supergravity theory.  Charged black holes with
  $\alpha=1$ (arising from  the string
theory) and  $\alpha=1/ 3$ (arising  from the $5d$ Kaluza-Klein
theory \cite{GA}), however,
correspond to solutions of $N=4$ and $N=8$ supergravity Lagrangians,
respectively. The Bogomol'nyi bounds
for charged dilatonic
black holes were derived  for  $\alpha
=1$ and $\alpha=1/3$  in  Refs.
\cite{KLOPP}  and \cite{GP},
respectively.}
  They correspond to the  spherically symmetric solutions of the
Lagrangian (\ref{lag})  with the  matter fields $T$ turned off, {\it i.e.},
$V\equiv
0$, however, with non-zero gauge fields $F_{\mu\nu}\ne 0$.
The  (Einstein frame) metric  is of the form
\cite{GM,HW}:\footnote{
The corresponding electrically charged black holes have the same Einstein
frame metric, however, the dilaton solution is related to the corresponding
magnetic one by the  transformation $\phi\rightarrow -\phi$.}

\begin{equation}
ds^2 = \lambda(r)dt^2 - \lambda(r)^{-1}dr^2 - R(r)d\Omega_2^2,\label{metbh}
\end{equation}
 with:
\begin{equation}
\lambda(r)=\left(1-{r_0\over r}\right)^{{2\alpha\over {1+\alpha}}}\ ,\ \
R(r)=r^2
\left(1-{r_0\over r}\right)^{{2\over {1+\alpha}}}\ .\label{lam}
\end{equation}
 The dilaton field $\phi$ and the magnetic field are of the form:
\begin{equation}
{\rm e}^{2\phi/\sqrt\alpha}=\left(1-{r_0\over r}\right)^{-{2\over {1+\alpha}}}\
,\ \  F_{\theta\phi}= P\sin\theta \ .\label{dil}
\end{equation}
Here $P$ is the magnetic charge of the black hole,
$r_0^2=P^2({{1+\alpha}\over \alpha})$ and the mass $M$ of the black hole is
\begin{equation}
M^2= P^2 \left({{1+\alpha}\over \alpha}\right)
\end{equation}

Interestingly, the global space-time structure (and the related
thermal properties) of the extreme magnetically charged
dilatonic black holes  bear  striking similarities
to the one  of the corresponding  domain wall configurations, however, now the
role of $\alpha$ is inverted:

\begin{itemize}
\item
$\alpha=\infty$   corresponds to the  case, when the dilaton field does not
couple to the gauge fields. The solution therefore corresponds to the
extreme  Reissner-Nordstr\" om black hole, which has the time-like singularity
at $r=0$ and $r=r_0$ corresponds to the Cauchy horizon.  Its
  global space-time  structure (see Figure 1b) in the $(r,t)$ direction  is
{\it the same} as the one  of
the Type I supergravity  walls  in the $(z,t)$ direction. In the latter case,
however,
the  time-like singularity is replaced by the  wall. The corresponding
temperature of the black hole $T=0$.
\item
$\infty>\alpha>1$ corresponds to  solutions with  the curvature singularity at
$r=r_0$. Null radial geodesics
reach $r=r_0$ in  infinite time, {\it i.e.},
affine time $\tau\equiv\int^{r_0}\ dr/\lambda=\infty $.  Therefore $r=r_0$
corresponds to the  null singularity (see Figure 1b).
The corresponding temperature $T=0$.
\item
$\alpha=1$ corresponds to the stringy extreme magnetically charged
black hole  with  the null singularity at $r=r_0$
 (see Figure 2b), however,
 the temperature $T= M/8\pi$ is finite.
\item
$\alpha<1$, corresponds to solutions, where the  singularity at $r=r_0$  is
reached  by a null geodesics in a
{\it finite} (affine) time. Thus,  the singularity is naked
 (see Figure 3b) and the temperature $T$ is infinite.
\end{itemize}
Thus, extreme magnetically charged
stringy dilatonic black holes ($\alpha=1$,  $T=M/(8\pi)$)  serve as a
 dividing line \cite{GM} between
extreme charged dilatonic black holes ($\alpha>1$, $T=0$) with the
singularity covered by the horizon and those ($\alpha<1$, $T=\infty$)  with the
naked singularity.

 We would like to emphasize the  complementarity
between   the extreme dilatonic domain walls and extreme
magnetically charged   dilatonic black holes.  The global space-time in the
 $(t,z)$ slice  for  extreme walls with coupling $\alpha$
 is the same as the one  in the $(t,r)$ slice for  extreme magnatically
chaged black holes  with coupling $1/\alpha$ (See Figures 1-3).
Between the two solutions the
role of   $\alpha$ is
inverted, while  the role of $W_M(T)$ on one side of the wall  and the
magnetic charge $P$
of the black hole are interchanged. The origin of such a complementarity can be
traced to the form of the Lagrangian (\ref{lag}), where  in the case of the
walls the potential (the source for the wall) is modulated by the dilaton
 coupling  of the type ${\rm
e}^{2\sqrt\alpha\phi}$, while in the case of the  black hole the kinetic
energy of the gauge field (the source for the magnetically  charged
configuration)  is modulated
by a complementary
 dilaton coupling of the type ${\rm
e}^{-2\phi/\sqrt\alpha}$.

One would further expect that a  two ($(t,z)$) dimensional
 effective action for the wall with  the metric  Ansatz (5), and a two
($(t,r)$) dimensional
effective action for
the black hole with the metric  Ansatz (10),
bear similiarities.\footnote{I would like to thank D. Youm for collaboration
on this point.}
 It, however, turns out that such a similarity between the
two actions is not transparent. One can, however, show
that near the singularity, the   metric (\ref{met}) (in the $(t,z)$)
slice) of the wall  with the coupling $\alpha$
 is the same as the metric (\ref{metbh})  (in the $(t,r)$ slice) of the  black
hole with the coupling $1/\alpha$. Namely, in the region
 $r-r_0\equiv\rho \rightarrow 0^+$, the
the coordinates ($t,\rho$)  of the black hole with the coupling
$\alpha$  and  the coordinates $(t,z)$ of the wall with the
coupling $\tilde \alpha\equiv1/\alpha$ are related  in the following way:
\begin{eqnarray}
[1-{1\over
2}(\tilde\alpha-1)\sigma_{ext} |z|]^{2\over{\tilde\alpha-1}}&=
({{\rho}\over{r_0}})^{2\alpha\over{(\alpha+1)}} \ , \ \alpha\ne 1\ \
\nonumber\\
{\rm e}^{-\sigma_{ext}|z|}&={{\rho}\over{r_0}}\ , \  \ \ \ \ \ \ \alpha=1\
\end{eqnarray}
 where
$r_0=2/[(1+\tilde\alpha)\sigma_{ext}]$
and $\tilde\alpha\equiv 1/\alpha$.

Near the singularity the
dilaton blows up  in both cases, however, unlike the  corresponding
two-dimensional
metric slices, the  coordinate dependence of
the dilaton near the singularity is {\it different} in either case.
This fact is also reflected in the different form of the
corresponding two-dimensional effective actions.

We have studied extreme domain walls in $N=1$ supergravity
with a general dilaton coupling   $\alpha$. We found that such
configurations are static, planar  configurations interpolating between
isolated supersymmetric vacua with the  varying dilaton field. Type I  walls
interpolate between the Minkowski vacuum and the supersymmetric vacuum with
the varying
dilaton. For $\alpha>1$ the walls have a (planar) naked singularity on one
side of  the wall, while for $\alpha\le 1$, the singularity is covered by the
horizon.  The  extreme magnetically charged black
holes  with the coupling $\alpha$ have  the same global space-time structure
 as the Type I wall with the coupling $1/\alpha$.
Interestingly, only in the  case with $\alpha=1$, which
corresponds to the
tree-level low-energy theory of $N=1$ $4d$ superstring vacua, the above types
of
configurations do not have naked singularities.
\acknowledgements

I would like to thank R. Myers, J. Russo,  and D. Youm for   useful
discussions.
The work is supported by  U.S. DOE  Grant No. DOE-EY-76-02-3071.

\vskip 150mm
{\bf Figure Captions}
\vskip 5mm
Figure 1:{ In Figure 1a  the Penrose-Carter diagram
in the $(z,t)$ direction  for
the extreme Type I ordinary domain wall ($\alpha=0$) is presented. It
corresponds to the most symmetric geodesic extension and comprises of
a  system of  an infinite lattice
of semi-infinite Minkowski space-times ($M$) separated by an anti-de Sitter
core.
The  compact null coordinates
define the axes: $u,v = 2\tan^{-1}(t \mp z)$.
The domain  wall region
is denoted with the thin lines.
Cauchy horizons (dashed lines) are the nulls separating
the anti-de Sitter  patches. Figure 1b represent the Penrose-Carter diagram
for  the extreme magnetically charged Reissner-N\" ordstrom black hole
($\alpha=\infty$) in the $(r,t)$ direction. The jagged line represents the
time-like singularity and the dashed lines are the corresponding Cauchy
horizons.  Note a formal similarity between Figures 1a and 1b.
 }
\vskip 5mm
Figure 2:{ In Figure 2a  the Penrose-Carter diagram in the $(z,t)$
plane for  extreme  Type I dilatonic domain wall  with $0<\alpha\le 1$ is
presented.  The compactified null coordinates  are  $u,v = 2\tan^{-1}(t \mp
z)$.
The wall (denoted by a thin line) separates the semi-infinite Minkowski
space-time ($M$) and the space-time
with a varying dilaton field and  the null singularity
covered by the horizon (jagged line).
Figure 1b corresponds to the Penrose-Carter diagram in
the $(r,t)$
plane for the extreme  magnetically  charged black hole $\infty>\alpha\ge 1$.
The jagged line corresponds to  the null singularity
covered by the horizon.}
\vskip 5mm
Figure 3:{ In Figure 3a  the Penrose-Carter diagram in the $(z,t)$ plane for
 extreme  Type I dilatonic domain wall  with $\alpha>1 $ corresponds to the
the wall (denoted by a thin line) separating the semi-infinite Minkowski
space-time ($M$) and the space-time with a varying dilaton field and the
naked (planar) singularity (jagged line). The
compactified null coordinates are $u,v = 2\tan^{-1}[t \mp (z+|z|_{sing})]$.
Figure 2b represents  the Penrose-Carter diagram for the
 extreme  magnetically charged dilatonic black hole ($\alpha<1 $)  in the
$(r,t)$ plane.  The   singularity (jagged line) is naked.}
\vskip 100mm
\begin{figure}[p]
\iffiginclude
\psfig{figure=marseille21.eps,height=110mm
}
\fi
\end{figure}
\vskip 100mm
\begin{figure}[p]
\iffiginclude
\psfig{figure=marseille23.eps,height=75mm
}
\fi
\end{figure}
\vskip 100mm
\begin{figure}[p]
\iffiginclude
\hbox{\hfill\psfig{figure=marseille22.eps,height=75mm
}}
\fi
\end{figure}
\end{document}